# Impacts of extreme weather events on terrestrial carbon sequestration revealed by weather stations in the Northern Hemisphere

Haiyang Shi[1,2]

[1] Department of Civil and Environmental Engineering, University of Illinois at Urbana-Champaign, Urbana, IL 61801, USA
[2] State Key Laboratory of Ecological Security and Sustainable Development in Arid Region, Chinese Academy of Sciences, Urumqi, 830011, China

Correspondence: Haiyang Shi (hiayang@illinois.edu; haiyang.shi9473@outlook.com)

**Abstract**

The increasing frequency of global climate extremes has significantly impacted the terrestrial carbon cycle. Extreme weather events such as heatwaves, droughts, and extreme precipitation pose serious threats to ecosystem carbon sequestration. This study investigated the impacts of these extreme events on terrestrial carbon sequestration using data from weather stations in the Northern Hemisphere, by combining weather station observations and machine learning-based Gross Primary Production (GPP) and ecosystem respiration (Reco) estimates. Droughts and heatwaves have the most significant impact, often correlated as compound events. The effects of extreme precipitation and cold extremes may have been underestimated in the past. Whether various extreme events occur in spring or summer led to different mechanisms. We provided a more precise and station-specific analysis compared to using coarse-resolution climate reanalysis and model simulations. It also suggests the need for improved methodologies and the integration of data-driven and process-based models to better predict and understand the effects of extreme weather on ecosystem carbon cycling.

**Introduction**

More frequent climate extremes[1] across the globe have seriously affected the terrestrial carbon cycle[2–7]. Extreme heatwaves[7–10], extreme droughts[2,7,11–13], extreme precipitation[14–17], cold extremes[18,19], etc., and their accompanying wildfires[20] (by droughts), soil erosion[21] (by extreme precipitation), etc., have been found to have serious negative impacts on ecosystem carbon sequestration. An in-depth assessment of the impacts of climate extremes on the carbon cycle is important for adaptation to global climate change, especially in the context of the recent increase in the frequency and intensity of climate extremes[1].

Extreme heatwaves[7–10], extreme droughts[2,7,11–13], extreme precipitation[14–17], and cold extremes[18,19] are common extreme climatic events that impact carbon fluxes, such as Gross Primary Production (GPP) and ecosystem respiration (Reco), potentially turning ecosystems from carbon sinks into carbon sources[11,12]. These impacts have been widely studied in previous studies:

a) The frequency and intensity of extreme precipitation may increase[22] in the context of global warming. In arid regions, extreme precipitation can enhance soil moisture, benefiting ecosystems. However, in humid regions, extreme precipitation can lead to flooding, causing root hypoxia, reduced respiration, and reduced vegetation growth[17,23]. Additionally, extreme precipitation can cause soil erosion, and soil carbon loss[24] and decrease nutrient levels.

b) Drought is often considered to negatively impact ecosystem carbon sequestration[2,7,11–13]. For instance, the carbon released during the European summer drought of 2003[7] and the Amazon droughts of 2005 and 2010[12,13] may have offset several years of carbon sequestration. Various process-based ecosystem models have been used to simulate the effects of drought[25], but there is significant variability and uncertainty in results from various models[25]. Previous studies have shown that drought typically negatively affects vegetation growth and GPP through increasing atmospheric dryness[26] and decreasing soil moisture[27]. Drought also affects soil respiration by reducing soil moisture[28] and may lower Reco by decreasing vegetation growth and autotrophic respiration. However, it can also increase soil respiration by increasing soil temperatures[29]. The mechanisms involved are not yet fully understood. Moreover, drought can lead to plant mortality[30] and indirectly increase wildfires[20] and desertification, which negatively impact ecosystems.

c) Warm extremes and heatwaves have become more frequent[31], often reducing photosynthesis in vegetation and potentially increasing Reco[29,32]. Warm extremes are frequently correlated with drought[33], forming compound heatwave-drought events[9,34], which can offset the positive effects of warm extremes on Reco with the negative effects of drought[28]. The season during which warm extremes occur also leads to different impacts; for example, warm extremes in spring can benefit vegetation growth[35] but may indirectly reduce summer water availability by increasing evapotranspiration in spring and thus negatively affecting summer vegetation growth[36].

d) In the Northern Hemisphere, due to warming, spring phenology has advanced, and autumn phenology has been delayed[37]. Therefore, during early and late growing seasons, cold extremes may more frequently affect ecosystems. Cold extremes can slow down vegetation growth[38] and even lead to plant mortality by frost damage. In agricultural regions, spring cold extremes can delay sowing and reduce crop yields[39,40].

However, previous studies have not yet accurately captured the impacts of extreme climate events on carbon sequestration capacity across the globe[3]. Observation-based methods such as eddy-covariance (EC) flux stations, while providing a relationship between measured carbon fluxes and climate extremes[41], are sparsely distributed globally and provide insufficient temporal coverage of extreme climate events. While process-based models[25] are also not yet able to accurately represent the impacts of climate extremes in their equations, data-driven approaches are becoming increasingly popular, which can upgrade carbon fluxes from EC station networks to the global scale by combining climate reanalysis data and remote sensing data[42]. However, climate reanalysis datasets often have non-negligible shortcomings in capturing extreme climate events[43], due to their dependence on climate models, which are not yet able to perfectly simulate extreme weather events.

Compared to reanalysis data, weather station observations can be more effective in characterizing climate extremes (e.g., extreme precipitation). Recent studies[44] have also demonstrated the feasibility of modeling carbon fluxes within a few hundred meters around weather stations. Therefore, this study aimed at combining weather station-based identification of extreme events and weather station-level GPP and Reco estimates to revisit the impact of climate extremes on ecosystem carbon sequestration. It is promising to efficiently assess the impact of climate extremes at the station scale and to provide more coverage of climate extremes in areas and periods not covered by flux station networks. We first used machine learning to estimate GPP and Reco at the weather station scale and identify multiple extreme climate events from station observations. We then analyzed the impacts of the various climate extremes on the current season and subsequent seasons and years. This study can update our understanding of the climate extreme-carbon sequestration capacity relationship.

**Extreme weather identification**

Using the 99.9th percentile of the 7-day mean of long-term climate observations from the stations, the thresholds and frequencies of four types of extreme events including extreme precipitation, warm extremes (characterized by daily maximum temperature), drought (represented by high VPD indicating atmospheric dryness), and cold extremes (characterized by daily minimum temperature) were identified (Fig. 1). These events exhibited significant geographical variability, with warm extremes and extreme droughts appearing to occur more frequently. This may be related to the correlation between the two, and global warming has also made it more likely for warm extremes to be identified in the last 20 years compared to historical periods.

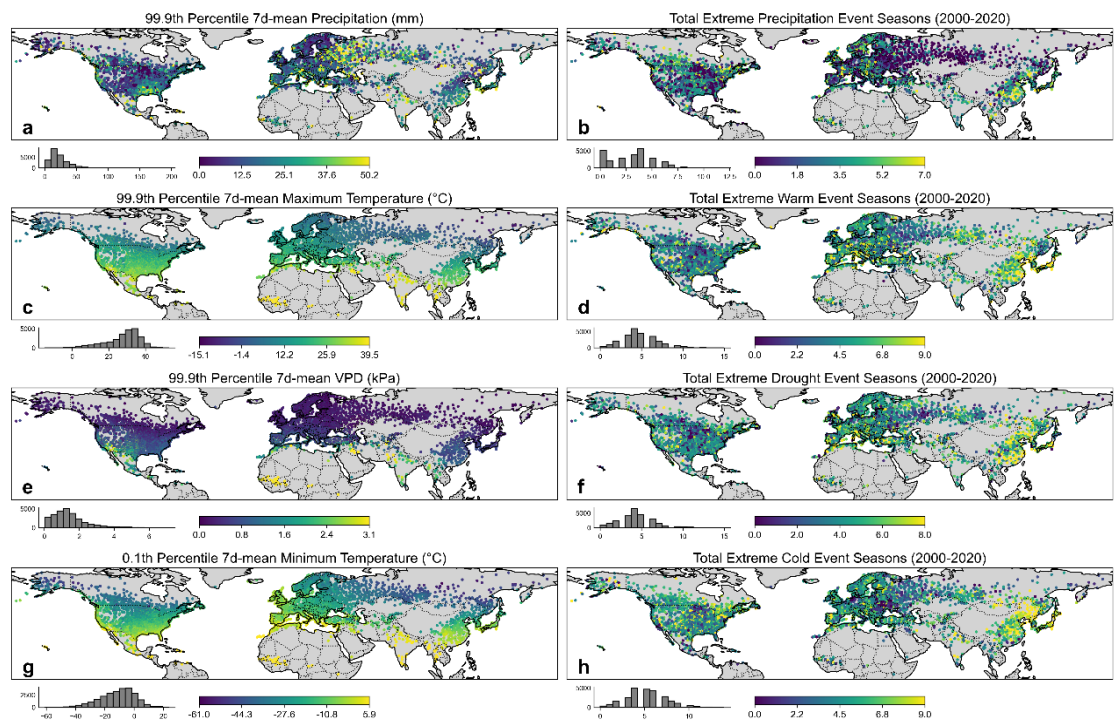

**Fig. 1** Thresholds for extreme precipitation, warm extreme, extreme droughts, and cold extremes (a,c,e,g)

versus their respective number of seasons of occurrence from 2000 to 2020 (b,d,f,h). Thresholds for extreme precipitation, extreme heat, and extreme drought are the 99.9th percentile of the 7-day moving mean of long-term observation records. The threshold for cold extremes is the 0.1th percentile of the 7-day moving mean.

The thresholds for extreme climate events observed at the stations indeed exhibited inconsistencies with the reanalysis data (Fig. 2). The 99th percentile of daily TAmax observed at the stations from 2016 to 2019 was generally higher than the values from ERA5-land, reaching 2 to 3 degrees Celsius higher at many stations. The 99th percentile of VPD observed at the stations also showed considerable differences from ERA5-land values; for example, at most stations in the United States, the observed VPD was 0.3 to 0.4 kPa lower than the ERA5-land values. Similarly, the 99th percentile of precipitation observed at the stations was generally higher than the ERA5-land values, especially in regions such as the East Asian monsoon area. This highlights the importance of using observed station data to assess extreme events.

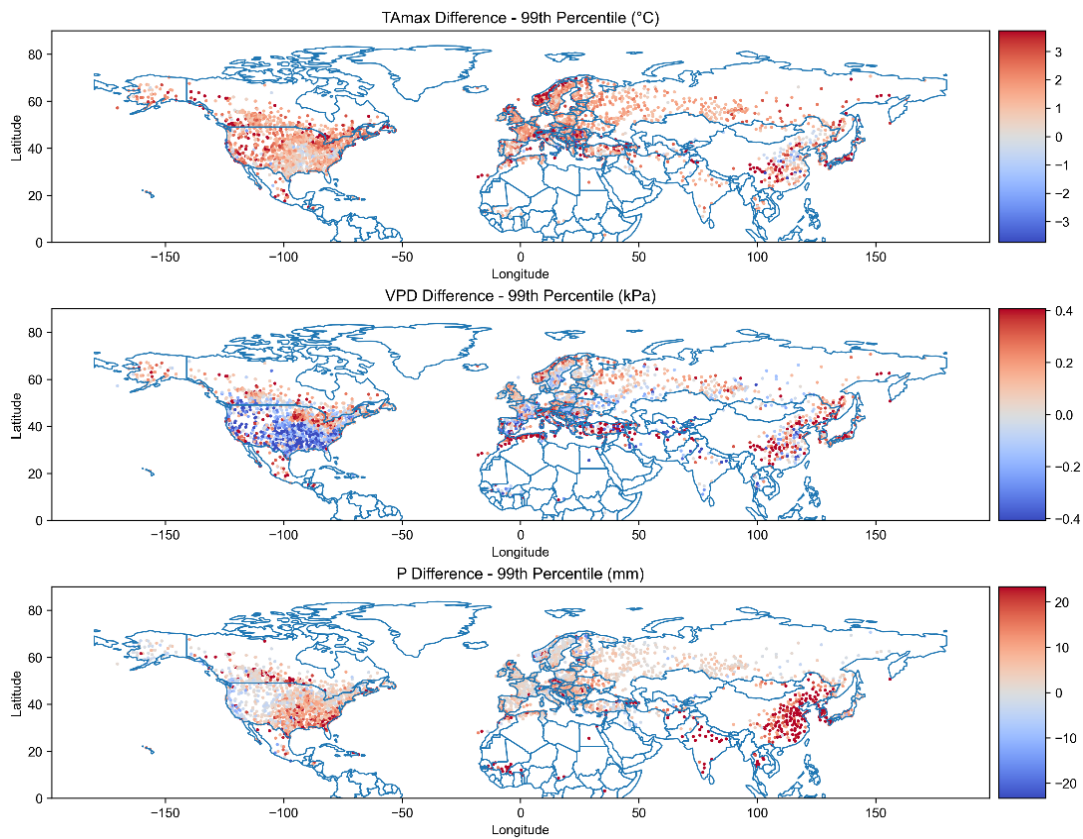

**Fig. 2** Differences (observation minus ERA5-land data) between the extremes (99th percentile) of the weather station's TAmax, VPD, and P and the extremes (99th percentile) extracted from the coarse-resolution ERA5-land reanalysis climate products from 2016 to 2019.

**GPP and Reco patterns**

The annual mean estimates of GPP and Reco from the meteorological stations are generally reasonable (Fig. 3). The magnitude of the negative anomalies in annual GPP may be higher than the positive anomalies in annual Reco, particularly in regions such as Europe. This indicates that it is more likely that extreme negative anomalies in GPP, rather than extreme positive anomalies in Reco, cause the ecosystem to become a carbon source. The coefficients of variation (CV) for GPP and Reco appear to be lower in regions with higher baseline values, and the spatial patterns are consistent. However, the CV for GPP in Europe seems to be greater than the CV for Reco, indicating greater interannual variability in GPP.

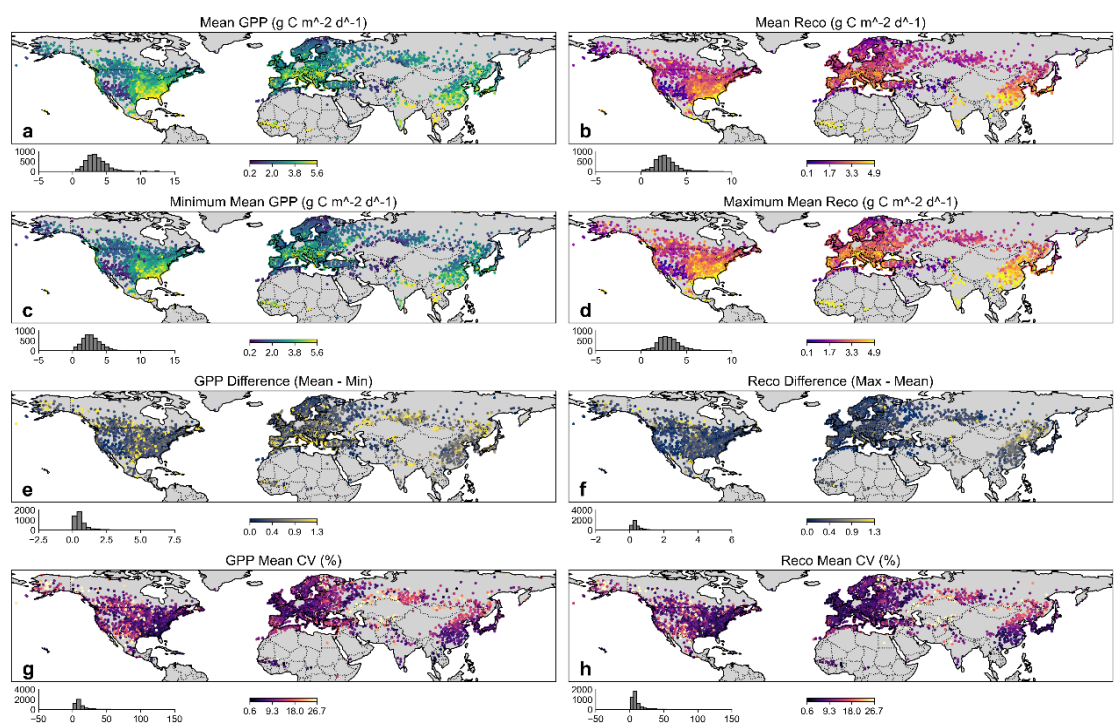

**Fig. 3** Mean (a-b), min/max (c-d), the difference from the mean (e-f), and coefficient of variation (g-h) of GPP and Reco at weather stations estimated by machine learning.

**Extreme precipitation**

We analyzed the impact of extreme precipitation events by season: When extreme precipitation occurs in spring, GPP and Reco tend to exhibit negative anomalies. This effect seems to be minimal in the summer. In arid regions, spring Reco is more likely to show positive anomalies under spring extreme precipitation compared to humid regions. This may be due to the trigger effect of precipitation events on soil respiration in the early growing season in arid areas[45]. In the event of extreme summer precipitation, both summer GPP and Reco exhibit small negative anomalies, especially in humid regions. This could be because extreme precipitation in humid areas can lead to prolonged flooding, oxygen deficiency in plant roots, reduced respiration, and decreased plant growth[15,17,23]. This effect appears to extend into autumn. When extreme precipitation occurs in autumn, both autumn GPP and Reco exhibit small negative anomalies. The mechanism may be similar to that in summer but could also be influenced by residual effects from summer (given that autumn extreme precipitation may correlate with high summer precipitation in wet years). The impact of extreme precipitation in each season on GPP and Reco in the following year appears to be small. However, positive anomalies in GPP and Reco can be found for the following summer in the arid areas. This could be the result of extreme precipitation recharging the water storage, allowing water to be supplied for plant growth in the following year.

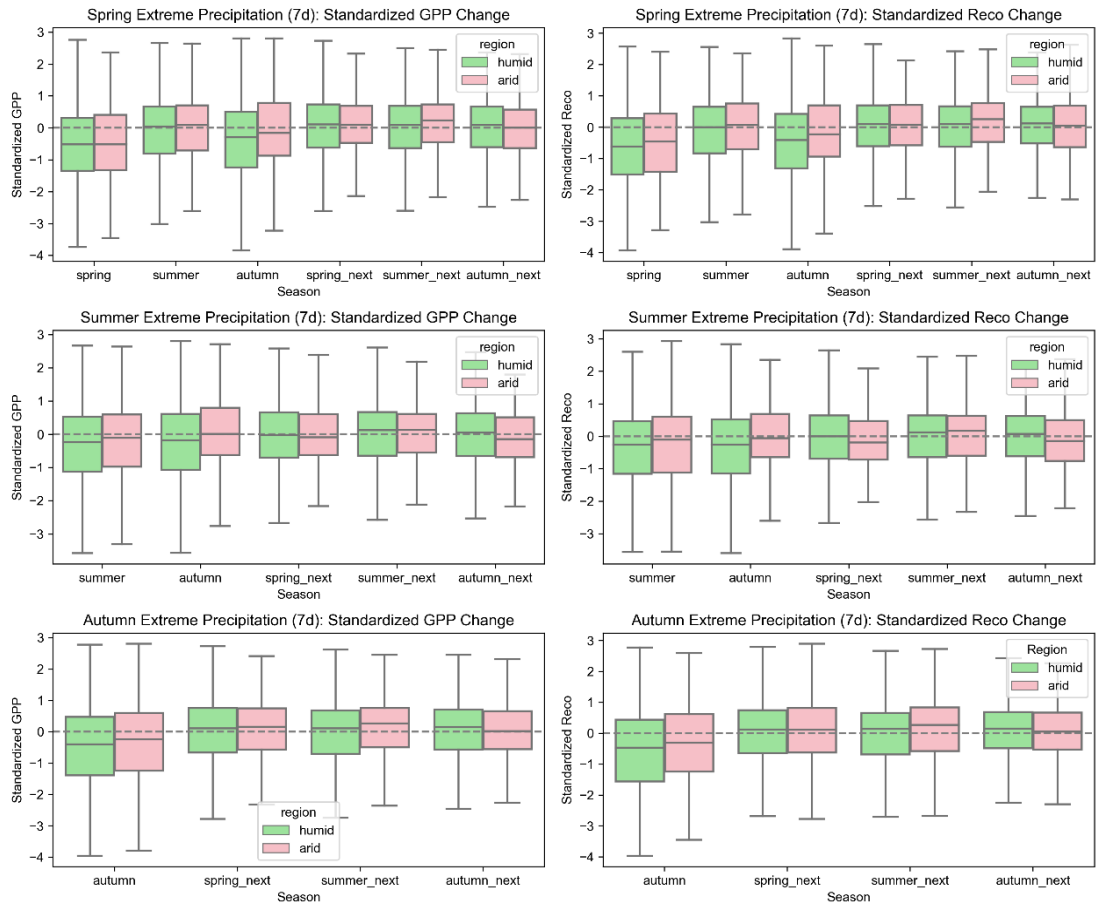

**Fig. 4** Anomalies in GPP and Reco in spring, summer, autumn, and the following year (next spring, summer and autumn) when spring precipitation extremes occur (upper part). Anomalies in GPP and Reco in summer, autumn, and the following year when summer precipitation extremes occur (middle part). Anomalies in GPP and Reco in autumn and the following year when autumn precipitation extremes occur (lower part). The five horizontal lines from top to bottom of each box indicate the 100, 75, 50, 25, and 0 quartiles of the data.

The negative effects of spring precipitation extremes on spring GPP and Reco are widespread (Fig. 5a & Fig. 5b). Positive effects of spring and summer precipitation extremes on summer GPP can be found at many stations in agricultural regions such as the Corn Belt in the central U.S. (Fig. 5c & Fig. 5e), where carry-over water appears to reduce the risk of crop water shortages. However, in regions such as eastern China and the southwestern U.S., summer precipitation extremes more broadly negatively affect summer GPP. This may be due to the higher intensity of precipitation extremes and accompanying storms, which can cause loss of branches and leaves, and increase tree mortality[3].

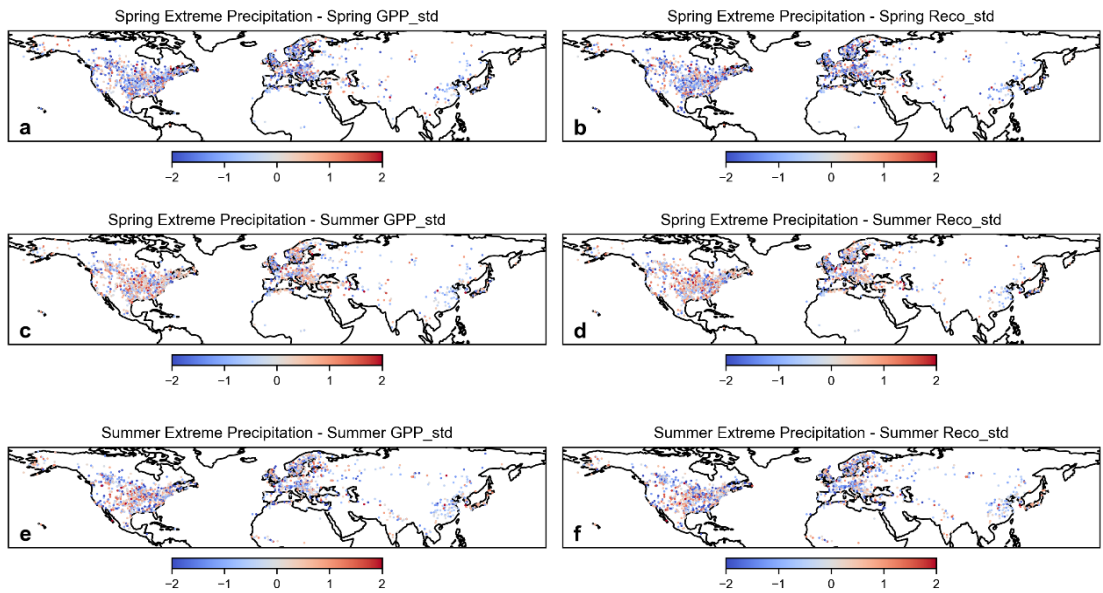

**Fig. 5** Anomalies in spring and summer GPP and Reco when spring precipitation extremes occur (a-d) and summer GPP and Reco when summer precipitation extremes occur (e-f). GPP_std and Reco_std denote the anomalies of the standardized GPP and Reco, respectively.

**Warm Extremes and Drought**

Under natural conditions, there is often a correlation between warm extremes and drought[9,46]. Warm extremes are often accompanied by decreased precipitation and lower relative humidity, which introduces uncertainty in accurately understanding the impacts of warm extremes. This is also related to the use of VPD extremes in this study to measure drought. Although VPD largely reflects atmospheric dryness[47], it is also influenced by temperature. Mechanistically, warm extremes often coincide with drought, such as the heatwaves in Europe that accelerated soil drying and warming[48].

The impact of warm extremes in spring is smaller than that in summer (Fig. 6), and their effect on summer GPP and Reco is reduced. Summer warm extremes have significant negative effects on both GPP and Reco. This can be partly attributed to drought (Fig. 7) due to their correlation. The magnitude of drought impacts appears to be slightly greater than that of warm extremes. This indicates that the combined effects of warm and dry events are substantial, which has been focused more frequently recently. The negative impacts of summer drought on GPP and Reco are greater in humid regions than in arid regions. By autumn, the residual effects of summer drought diminish, and its lagged effects on the following year appear to be limited.

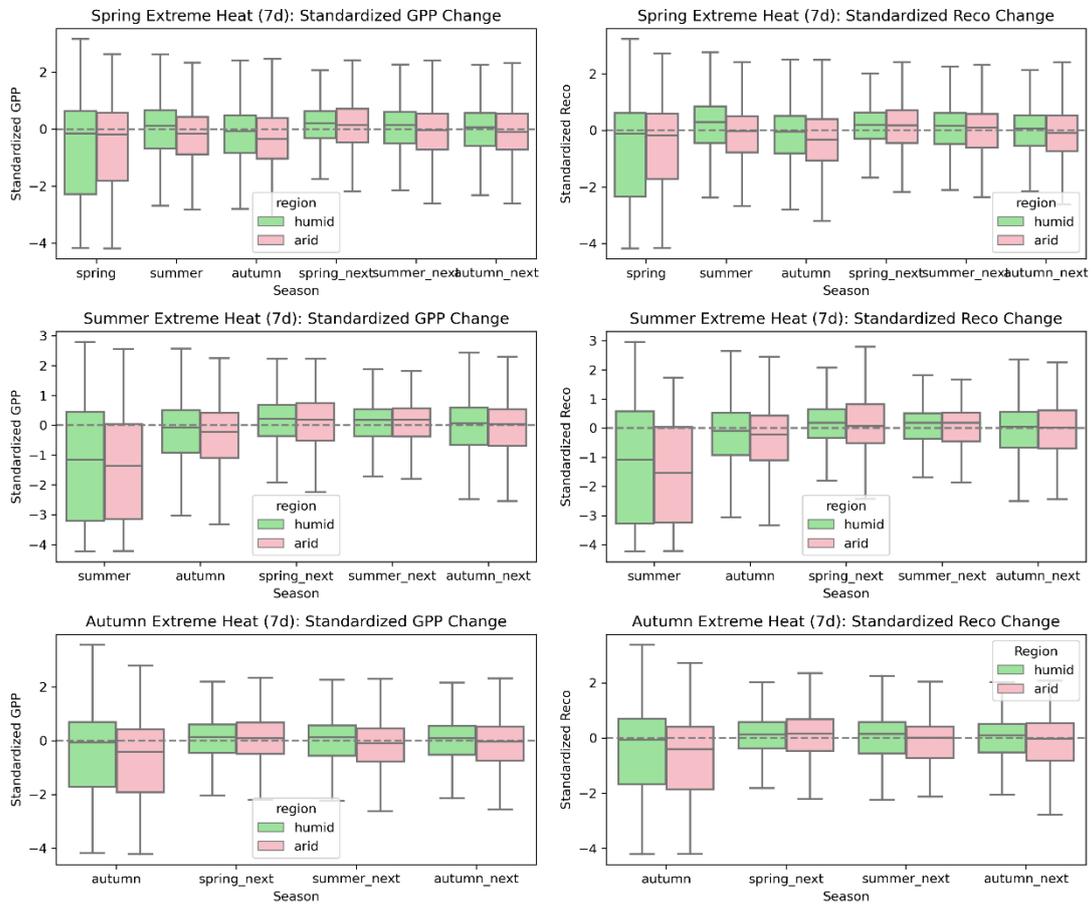

**Fig. 6** Anomalies in GPP and Reco in spring, summer, autumn, and the following year (next spring, summer and autumn) when spring warm extremes occur (upper part). Anomalies in GPP and Reco in summer, autumn, and the following year when summer warm extremes occur (middle part). Anomalies in GPP and Reco in autumn and the following year when autumn warm extremes occur (lower part). The five horizontal lines from top to bottom of each box indicate the 100, 75, 50, 25, and 0 quartiles of the data.

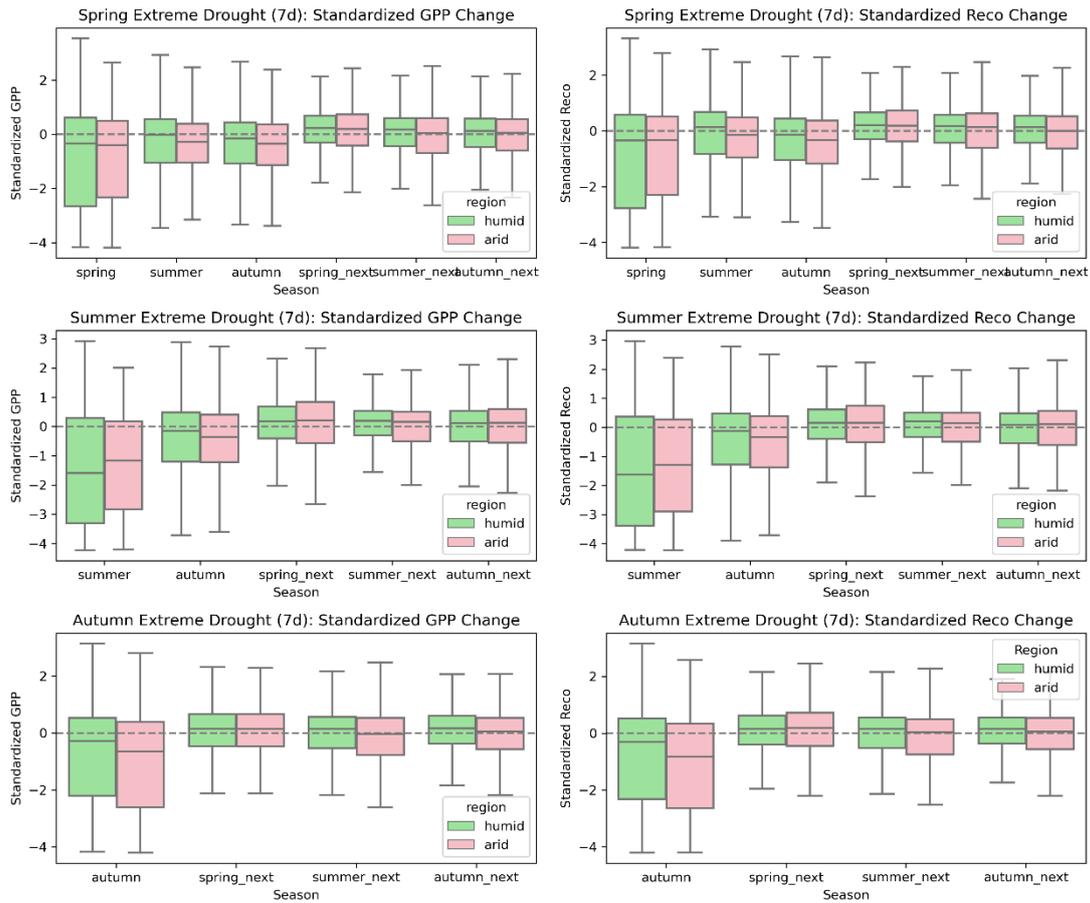

**Fig. 7** Anomalies in GPP and Reco in spring, summer, autumn, and the following year (next spring, summer and autumn) when spring drought occurs (upper part). Anomalies in GPP and Reco in summer, autumn, and the following year when summer drought occurs (middle part). Anomalies in GPP and Reco in autumn and the following year when autumn drought occurs (lower part). The five horizontal lines from top to bottom of each box indicate the 100, 75, 50, 25, and 0 quartiles of the data.

The spatial patterns of the impacts of warm extremes (Fig. 8) and drought (Fig. 9) also show similarities. However, differences remain. In some regions, such as southeastern China, the central U.S., and eastern Europe, warm extremes in spring may accelerate vegetation growth to some extent, leading to higher GPP and Reco in summer, provided that water deficits are not severe. However, the impacts of spring drought on summer GPP and Reco show that some stations, particularly many stations in the central U.S., shift from positive to negative impacts, especially for Reco. The negative impacts of drought on Reco may exceed the positive impacts of warm extremes. Spring drought can reduce the water carried over to summer, leading to soil moisture deficits[36]. This can decrease soil respiration and reduce leaf respiration due to leaf area decrease (i.e., leaf drop). Humid regions are more sensitive to these impacts (Fig. 7) because the possible greater soil moisture decline and higher dependence on surface soil moisture result in a more significant impact compared to arid regions. Geographic differences may also relate to the extent to which the warm extreme thresholds at each station exceed the optimal temperature[49] for Reco, although this may be partially confounded by drought.

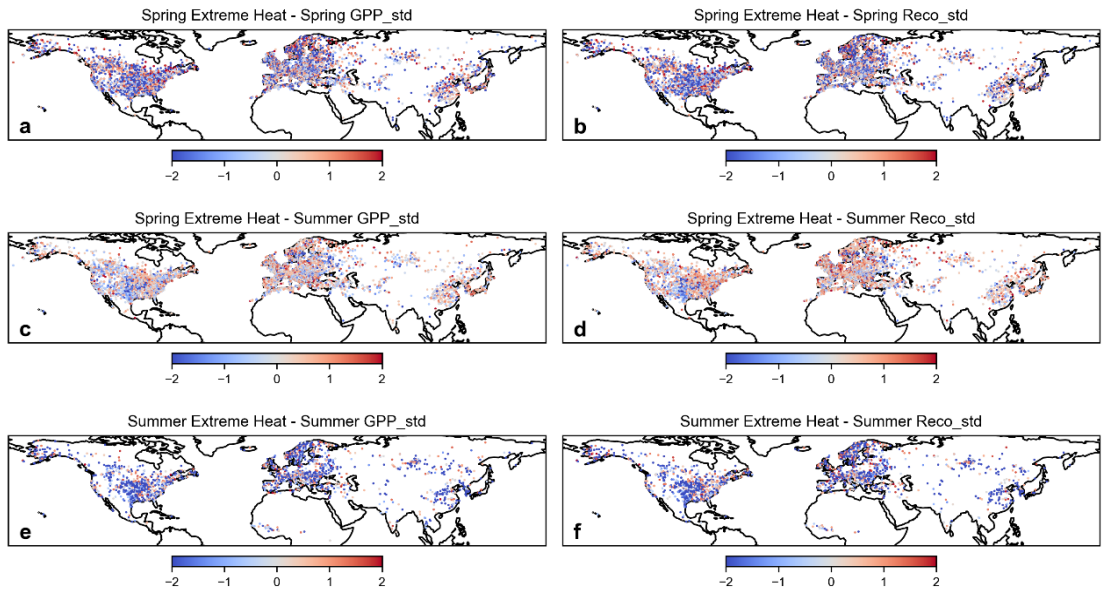

**Fig. 8** Anomalies in spring and summer GPP and Reco when spring warm extremes occur (a-d) and summer GPP and Reco when summer warm extremes occur (e-f).

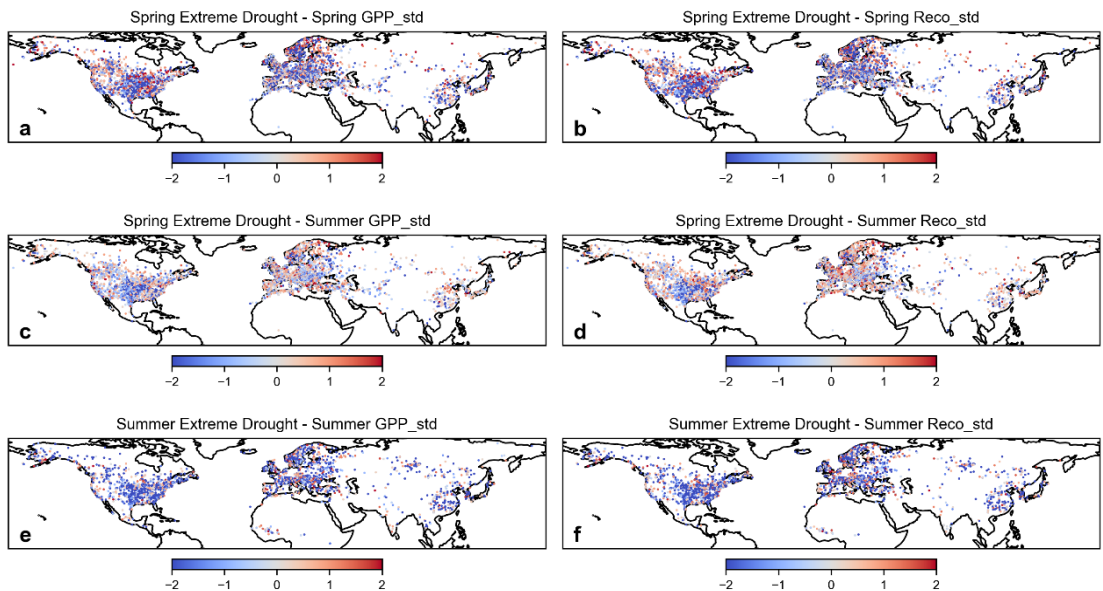

**Fig. 9** Anomalies in spring and summer GPP and Reco when spring drought occurs (a-d) and summer GPP and Reco when summer drought occurs (e-f).

**Cold Extremes**

Spring cold extremes have significant negative impacts on both spring GPP and Reco (Fig. 10). These effects are observed across most regions in the Northern Hemisphere (Fig. 11). Autumn cold extremes also cause slight negative impacts on autumn GPP and Reco. Compared to spring and autumn, summer cold extremes are less likely to be caused by frost, thus their impact is not as widespread as in spring. However, they negatively affect GPP and Reco in high-latitude regions (such as Northern Europe, Siberia, and Northern North America). This is because temperature is a critical variable in energy-limited areas, and cold extremes can occur alongside insufficient sunlight, further negatively impacting GPP.

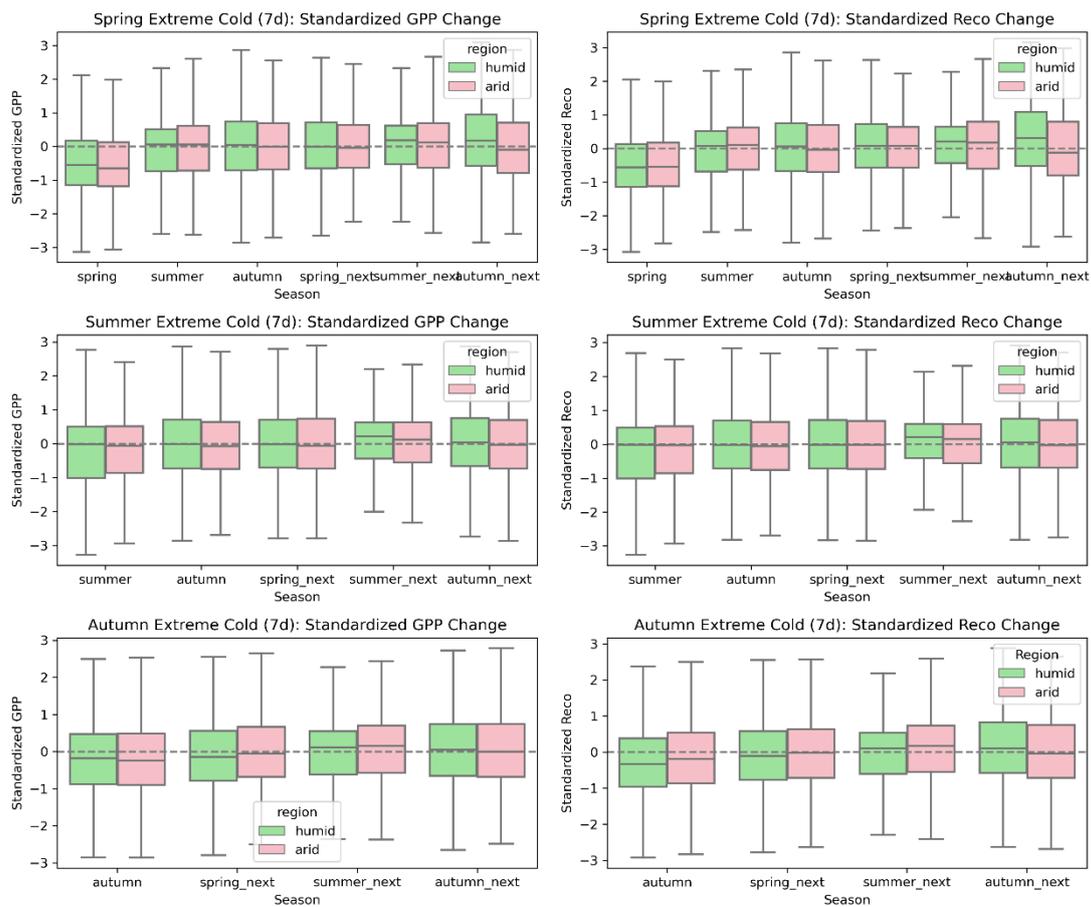

**Fig. 10** Anomalies in GPP and Reco in spring, summer, autumn, and the following year (next spring, summer and autumn) when spring cold extremes occur (upper part). Anomalies in GPP and Reco in summer, autumn, and the following year when summer cold extremes occur (middle part). Anomalies in GPP and Reco in autumn and the following year when autumn cold extremes occur (lower part). The five horizontal lines from top to bottom of each box indicate the 100, 75, 50, 25, and 0 quartiles of the data.

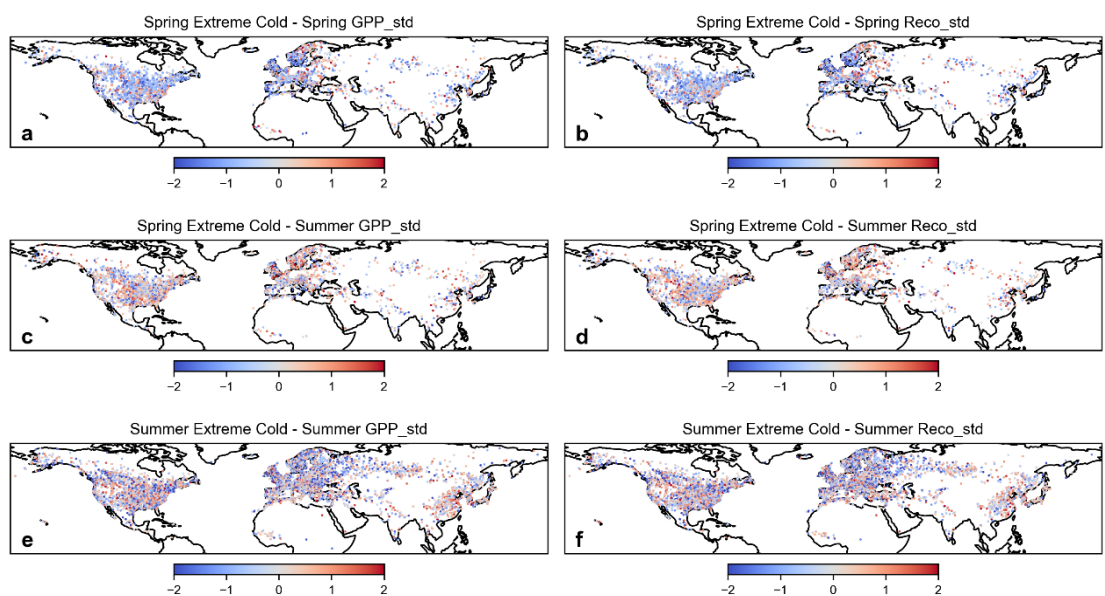

**Fig. 11** Anomalies in spring and summer GPP and Reco when spring cold extremes occur (a-d) and summer GPP and Reco when summer cold extremes occur (e-f).

**Extreme event type with the average greatest impact**

We compared the average magnitude of the impacts of different types of spring extreme events to identify the type of extreme event with the greatest impact at each site (Fig. 12). Drought and heatwaves ranked first and second, respectively, in terms of the average impact of spring extreme events. Particularly in the US and Europe, the type of event with the greatest average impact at most sites can be categorized as either drought or heatwaves. Given the correlation between the two, combining the number of sites affected by them may reveal a larger impact range of compound heat-drought events. Compared to GPP, warm extremes rank higher in types affecting Reco, indicating a higher sensitivity of Reco to warm extremes.

When comparing the impact magnitudes of different types of extreme events, the thresholds used to identify these extreme events may affect the comparability between different events. For instance, even when using the same percentile, the actual severity of the extreme events represented by the 99.9th percentile of precipitation and the 99.9th percentile of warm extremes can differ. For example, in arid regions, the 99.9th percentile of precipitation may not correspond to an extremely high absolute rainfall amount (compared to the 99.9th percentile of precipitation in humid regions) and may not even cause flooding. This could introduce uncertainty in ranking the impacts of the four types of events at different sites. Previous studies[3,50] have found that most of the 100 most severe negative fraction of absorbed photosynthetically active radiation (fAPAR) anomalies from 1982 to 2011 can be explained by droughts or warm extremes, but some events are not caused by warm extremes or droughts and may be due to events such as large-scale wind throw or pest outbreaks. Similarly, our results do not deny that GPP and Reco anomalies at some of these sites may not be caused by one of the four events singularly, but may be caused by other factors interfering or by complex responses to multivariate extreme events. Including land cover/use change, irrigation, agricultural land management, and nitrogen deposition may interfere with the pure quantification of the average magnitudes of various extreme event impacts.

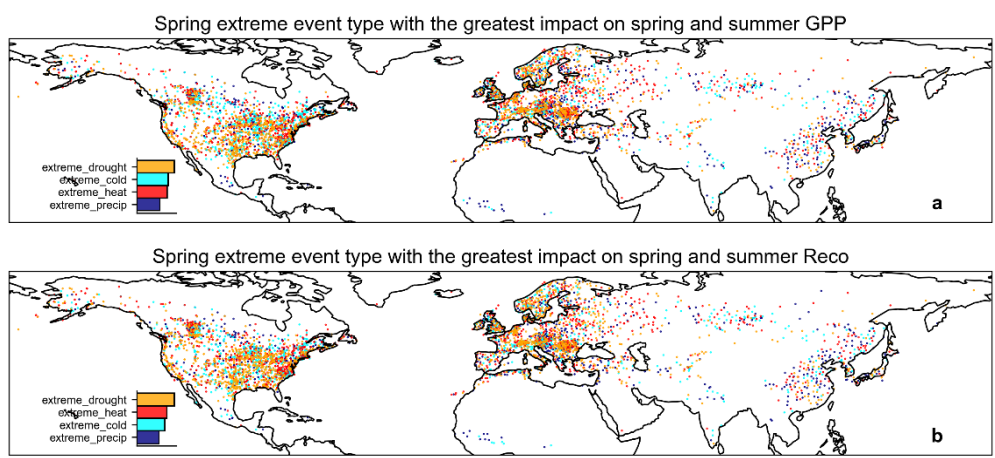

**Fig. 12** Distribution of the type of extreme spring event that has the average greatest impact on (a) GPP and (b) Reco of spring and summer.

**Discussion**

This study combined various extreme events detected by weather stations with GPP and Reco estimated at the station scale using machine learning to analyze the impact of extreme events on terrestrial carbon sequestration over the past 20 years. Compared to the coarse-scale, high-uncertainty datasets simulated using climate reanalysis data and ecological process models[25], the data we used are more station-specific and have higher precision. This approach can limit the uncertainty in assessments, thereby more accurately quantifying the impact of extreme events, which is very sensitive to the spatiotemporal resolution and accuracy of the data.

Droughts and heatwaves have a significant impact among various extreme events, and we also found a correlation between them as well as differences (such as their different impacts on Reco). This study used extreme values of VPD to represent drought, which may be somewhat influenced by high temperatures, thus more likely reflecting compound heatwave-drought events[9,34]. It may be useful to use more representative indicators of drought[51], including atmospheric, ecological, and hydrological drought indices, such as the Standardized Precipitation-Evapotranspiration Index (SPEI), leaf area index (LAI), vegetation optical depth[52], soil moisture, and precipitation minus evapotranspiration (P-ET)[47]. This could provide a more comprehensive understanding. Compared to droughts or heatwaves, extreme precipitation and cold extremes have received less attention in previous studies. This study quantified the effects and shows that they also have a significant impact on GPP and Reco.

Our results are more representative of areas with dense human footprints in the Northern Hemisphere, such as Europe, North America, and East Asia, where there is a relatively dense distribution of weather stations. However, in remote areas or regions with limited availability of meteorological data, the representativeness may be insufficient. Another limitation is the relatively low number of forest weather stations. Forests usually play an important role in the assessment of terrestrial carbon sequestration capacity under extreme climate conditions. For instance, the response of the Amazon rainforest to extreme climate events has already received widespread attention, which is crucial for the overall terrestrial carbon balance. However, the number of forest weather stations globally is relatively limited. In the future, combining data-driven approaches with forest flux station and weather station observations, microclimate observations[53], and multi-source remote sensing data[52,54,55] may have potential.

Additionally, the methodology for characterizing the impacts of extreme events should be improved.

As a data-driven approach, machine learning methods still have the potential to underestimate values under extreme conditions[56], especially when the amount of extreme climate data constitutes only a small fraction of the total dataset. As observations of various extreme events accumulate, the collaborative development of data-driven and process-based models will be crucial, particularly in predicting the future impacts of extreme climates. By more comprehensively integrating advancements in both methodology and data, it is promising to achieve a more accurate understanding of the impacts of extreme climates on ecosystem carbon cycling in the future and enhance our ability to adapt to extreme climates.

**Methods**

Weather stations' daily GPP and Reco estimations are from a dataset[57] produced by our team, which combined the FLUXNET2015 dataset, GSOD global weather station's meteorological observations and remote sensing. The data is produced by a prediction model based on the Long Short Memory Network (LSTM). Predictor variables used included the station's daily mean air temperature, maximum and minimum air temperatures, precipitation, wind speed, VPD (calculated by dewpoint temperature and mean temperature), LAI, downward shortwave radiation, elevation, slope and soil texture. The model has high accuracy, and in the leave-one-station-out cross-validation, the R-squared of most flux stations exceeds 0.6. In addition, we evaluated the extrapolability of the model when applied to weather stations (applicability at each weather station) based on the principle of geographic similarity (the assumption that stations with similar environmental variables are likely to have similar prediction accuracies). An R-squared of higher than 0.5 can be achieved for more than 60% of the weather stations and most stations with high accuracy are located in the northern hemisphere. We used only those weather stations where the projected R-squared exceeded 0.5.

Stations with fewer than 5 years of observations were excluded, as were years with fewer than 270 days of observations at each station. The 99.9th percentile of observations at each station on a 7-day scale (using moving averages over a 7-day window) was used as a threshold for the identification of extreme events (including extreme precipitation, drought, and warm extremes). Cold extremes were identified using the 0.1 percentile of daily minimum temperatures as the threshold. The timing of these extreme events was grouped into seasons (March through May for spring, June through August for summer, and September through November for autumn) to determine if a particular extreme event occurred during those seasons at each station.

Daily GPP and Reco were similarly aggregated to seasonal scales when analyzing the effects of climate extremes on them. For each station's GPP and Reco for a given season, we normalized them (calculated as z-scores) using the multi-year GPP and Reco values of the season. We then separately identified anomalies in GPP and Reco for the seasons following each season in which an extreme event occurred. If the extreme event occurred in spring, we matched GPP and Reco for spring, summer, and autumn of this year, as well as spring, summer, and autumn of the following year. If the extreme event occurred in summer, we matched GPP and Reco for summer, and autumn of this year, as well as spring, summer, and autumn of the following year. If extreme weather occurs in the autumn, we match the GPP and Reco for the autumn of this year and the spring, summer, and autumn of the next year. We also analyzed what the average most impactful extreme event type was for each station. The average anomalies magnitudes of GPP and Reco corresponding to each extreme event type were compared. Extreme event type with the largest average anomalies was then identified as the type with the greatest impact at that station. This type of identification was implemented for spring and summer, with spring extreme events assessed by their average impacts over the spring and summer of the year, and summer extreme events assessed by their impacts over the summer of the year. Differences in the impacts of extreme events were also differentiated between arid and humid areas. Areas with an Arid index lower than 0.5 were categorized as arid and those higher than 0.5 as humid.

To compare the ability of station observations and climate reanalysis data to characterize extreme climate events, we extracted daily precipitation, maximum air temperature, and VPD data of ERA5-land from 2016 to 2019. After matching them to weather station observations' dates, we calculated the difference between the 99th quantile values represented by the ERA5-land data and the 99th quantile values calculated from weather station observations.

**Acknowledgement**

This research was supported by the Dashboard for Agricultural Water use and Nutrient management (DAWN) project (Grant 2020-68012-31674) from the U.S. Department of Agriculture National Institute of Food and Agriculture and the Key projects of the Natural Science Foundation of Xinjiang Autonomous Region (Grant No. 2022D01D01).